\documentclass{aa}  

\usepackage{graphicx}
\usepackage{txfonts}
\usepackage[colorlinks=true,linkcolor=blue,citecolor=blue,urlcolor=blue]{hyperref}
\usepackage{physics}
\usepackage{amsmath}
\usepackage{xfrac}
\usepackage{float}

\begin{document} 

   \title{Competition between gravity waves excited by convection\\ and tides in stars that host a companion}

   \author{M. Esseldeurs\inst{1} \and J. Ahuir\inst{2} \and L. Amard\inst{3} \and S. Mathis\inst{4} \and L. Decin\inst{1}}

   \institute{Instituut voor Sterrenkunde, KU Leuven, Celestijnenlaan 200D, 3001 Leuven, Belgium \\
              \email{mats.esseldeurs@kuleuven.be}
         \and
             CEA, DAM, DIF, F-91297 Arpajon, France
         \and
             Observatoire de Genève, Université de Genève, Chemin Pegasi 51, 1290, Sauverny, Switzerland
         \and
             Université Paris-Saclay, Université Paris Cité, CEA, CNRS, AIM, 91191 Gif-sur-Yvette, France
             }

   \date{Received 22 November 2025; accepted 27 March 2026}

  \abstract%
        {Asteroseismology has become a powerful diagnostic tool in stellar astrophysics, offering unprecedented insights into the internal structures and dynamics of stars. Through the analysis of stellar oscillation modes, it enables precise characterization of stellar interiors across a wide range of stellar masses and of evolutionary phases, from the main sequence to the white dwarf phase. At the same time, the number of detected close stellar and planetary companion discoveries throughout the entire stellar evolutionary phases has increased significantly, prompting key questions about the interplay between stellar evolution and binarity.}
        {In this study, we investigate the competition between gravity waves (IGW) excited by internal convection and those excited by tides in stars that host a companion. By modelling the energy and angular momentum luminosities transported by IGWs stochastically excited by convection and by tides, we seek to quantify their relative contributions and identify the key parameters that govern their efficiency.}
        {We compute the energy and angular momentum luminosities transported by both type of waves for a range of stellar masses and evolutionary stages, with a particular focus on understanding how the presence of a companion influences the angular momentum transport and the induced rotational evolution of the radiative layers of the host star.}
        {The competition between the two excitation mechanisms is sensitive to the mass and orbital properties of the companion, as well as the internal structure of the host star. We find that for a Jupiter-mass companion, the stochastic excitation dominates over tidal excitation during all evolutionary phases. Only for close-in stellar companions around late type stars does the tidal excitation become more efficient.}
        {The presence of a companion is unlikely to significantly alter the internal angular momentum transport in the radiative layers of the host star, simplifying the modelling of IGW-driven angular momentum transport in stars that host a companion.}


   \maketitle
%

\section{Introduction}
    Asteroseismology has emerged as one of the most powerful diagnostic tools in stellar astrophysics, offering unprecedented insights into the internal structure and dynamics of stars. By analysing stellar oscillation modes, it is possible to probe the otherwise inaccessible interior of stars and to constrain key processes such as rotation, mixing, and energy transport. Over the past two decades, asteroseismology has transformed our understanding of stellar evolution across a wide range of masses and evolutionary phases, from main-sequence (MS) stars to compact remnants such as white dwarfs (WDs) \citep{Aerts2010, Garcia2019,Christensen-Dalsgaard2021, Aerts2021,Bowman2026}.

    The advent of high-precision, long-duration space-based photometry from missions such as CoRoT \citep{Baglin2006}, Kepler \citep{Borucki2010}, and TESS \citep{Ricker2015} has been especially transformative. These facilities have enabled the direct measurement of internal rotation at different points inside of the star, revealing a consistent pattern of strong angular momentum redistribution from the stellar core to the envelope \citep{Beck2012,Mosser2012,Deheuvels2012,Deheuvels2014,VanReeth2016,Gehan2018,Li2020,Aerts2025}. Remarkably, the efficiency of this redistribution appears to be up to two orders of magnitude stronger than predicted by stellar evolution models taking into account a radial shellular differential rotation \citep{Zahn1992} and the related large-scale meridional circulation and hydrodynamical turbulence \citep{Eggenberger2012,Ceillier2013,Marques2013,Cantiello2014,Ouazzani2019,Aerts2019}. This discrepancy highlights a critical gap in our theoretical framework and has motivated intensive efforts to identify the physical mechanisms capable of transporting angular momentum with the required efficiency.

    Among the proposed candidates, the two main mechanisms are magnetic fields and internal gravity waves (IGWs). Winding-up of weak magnetic fields by differential rotation can generate strong toroidal fields, which can drive through their instabilities \citep[e.g.][]{Tayler1973} a Tayler-Spruit dynamo \citep{Spruit2002}. This may lead to efficient angular momentum transport via magnetic stresses \citep{Petitdemange2023,Barrere2026} invoked to explain the nearly uniform rotation of the solar radiative interior \citep{Eggenberger2005, Eggenberger2022a}, and core rotation rates of evolved low-mass stars \citep{Fuller2019, Eggenberger2022b}. IGWs are excited at convective-radiative interfaces by turbulent motions in the convective regions of stars \citep{Press1981,Goldreich1990,Schatzman1993,Belkacem2009,Lecoanet2013,Pincon2016} and can efficiently mix chemical species and transport angular momentum in the radiative interior. Early pioneering work on the role of IGWs in stellar evolution dates back to \citet{Press1981}, \citet{GarciaLopez1991}, \citet{Schatzman1993}, and \citet{Zahn1997}. These studies already demonstrated that IGWs may play a crucial role in shaping stellar rotation profiles. More recently, IGWs have been invoked to explain the nearly uniform internal rotation of the Sun \citep{Charbonnel2005} and solar-type stars \citep{Talon2005}. Their influence has also been further explored in more massive stars \citep{Rogers2013,Rogers2015}, as well as evolving subgiants \citep{Fuller2014,Pincon2017}.

    At the same time, observational surveys have revealed an ever-growing number of close stellar and planetary companions across virtually all evolutionary phases \citep[e.g.]{Sana2012,Mulders2018,Beck2022,Hinkel2024,Esseldeurs2026}. These discoveries have raised fundamental questions about the interplay between stellar evolution and binarity, for instance regarding the impact of companions on the internal and rotational evolution of their hosts \citep[e.g.][]{Mathis2018,Ahuir2021b}. One of the primary pathways through which a companion can alter the evolution of its host is via tides \citep{Zahn1994,Song2016}. The gravitational interaction between the two bodies generates an ellipsoidal deformation of the host star, commonly referred to as the tidal bulge (or equilibrium tide). Dissipation of this bulge in the convective envelope through turbulent viscosity gives rise to the so-called equilibrium tide \citep{Zahn1966a,Zahn1989a,Remus2012}, while excitation of waves (e.g. IGWs at the convective–radiative interface) constitutes the dynamical tide.

    Theoretical investigations of tidally-excited IGWs date back to the seminal works of \citet{Zahn1975,Zahn1977} and \cite{Goldreich1989}, who first outlined the physical framework for tidal IGWs dissipation in stars. Since then, the problem has been further developed in the context of wave excitation, propagation, and dissipation \citep[e.g.][]{Ogilvie2004,Ogilvie2007,Mathis2015,Barker2020,Ahuir2021a,Esseldeurs2024}. While much of this work has focused on their role in tidal rotational and orbital evolution, tidally excited IGWs are also carriers of energy and angular momentum within the star \citep{Goldreich1989,Talon1998}. This raises an intriguing question: to what extent do tidally induced gravity waves compete with, or even dominate over, the stochastic IGWs generated by convection?

    In this study, we investigate the competition between these two wave excitation mechanisms: stochastic excitation by convection and tidal excitation by a companion in stars across different evolutionary stages. Specifically, we model the associated energy and angular momentum luminosities and quantify their relative contributions under varying stellar and orbital conditions. We analyse a range of host star masses and evolutionary phases, with particular emphasis on how the presence of a close companion modifies the angular momentum transport of the host. We explore both stellar and planetary companions, examining how their masses and orbital separations influence the transport processes. In addition, we investigate the role of the host star's internal structure and its evolution in regulating the efficiency of these processes.

    The remainder of this paper is organised as follows. In Sect.~\ref{sec:framework}, we describe the theoretical framework used to model the energy and angular momentum transport by IGWs. In Sect.~\ref{sec:competition}, we analyse the competition between convective and tidal excitation mechanisms, focusing on the dependence of their relative importance on key stellar and orbital parameters. Finally, in Sect.~\ref{sec:discussion}, we discuss the broader implications of our results for stellar rotation. Finally a summary and conclusions are presented in Sect.~\ref{sec:conclusion}.

\section{Energy and angular momentum luminosities}\label{sec:framework}
    \subsection{Gravity waves in stars}
        Gravity waves (g-waves) are oscillations in a star that are restored by buoyancy. In stars, they propagate in stably stratified radiative zones, with frequencies lower than the Brunt-Väisälä frequency $N$ (in rad/s) given by \citep{Aerts2010}:
        \begin{equation}
            N^2=g_0\left(\frac{\partial_r p_0}{\Gamma_1 p_0}-\frac{\partial_r \rho_0}{\rho_0}\right)\ ,
        \end{equation}
        where $g_0$, $p_0$, and $\rho_0$ are the unperturbed gravitational acceleration (in cm/s$^2$), pressure (in g/cm s$^2$), and density (in g/cm$^3$), respectively. $\Gamma_1 = {(\partial \ln p_0 / \partial \ln \rho_0)}_S$ is the first adiabatic exponent, where $S$ is the macroscopic entropy.

        In stars, gravity waves can be excited by various processes. To become excited, they need a source of energy, which can be provided at the interface between the convective and radiative layers. This energy can be provided by stochastic processes, such as turbulent convection, or by tidal interactions with a companion. The computation of the energy and angular momentum luminosities of these processes will be described in the following sections.

        Once excited, gravity waves can propagate through the star. When propagating through the radiative zone, they can be dissipated by radiative damping. If this process is sufficiently strong, the waves will be dissipated before they reach the end of the radiative zone (either a convective zone, the center or the surface of the star). This is the regime of progressive waves. If the damping is weak, the waves will be able to propagate to the end of the radiative zone and reflect back inwards. In this case, the waves create standing modes. The transition between these two regimes is characterised by the critical frequency, $\omega_\mathrm{crit}$, separating g-modes and progressive gravity waves regimes. $\omega_\mathrm{crit}$ (in rad/s) is given by \citep{Alvan2015}
        \begin{equation}\label{eq:CriticalFrequency}
            \omega_\mathrm{crit} = {[l(l+1)]}^{\frac{3}{8}}{\left(\left| \int_{r_\text{in}}^{r_\text{out}} K_{\mathrm{T}} \frac{N^3}{r_1^3} \mathrm{~d} r_1\right|\right)}^{\frac{1}{4}}\ ,
        \end{equation}
        where $K_{\mathrm{T}}$ is the thermal diffusivity (in cm$^2$/s), $r_\text{in}$ and $r_\text{out}$ are the inner and outer radii of the radiative zone, respectively, and $l$ is the spherical harmonic degree of the wave. In this work we will only consider the case of progressive waves. Hence, we will only consider the case where the waves are excited with a frequency below the critical frequency. This is the most suitable regime for angular momentum transport, as low-frequency IGWs are those which are the most efficiently damped, which is necessary for angular momentum to be transported \citep[e.g.][]{Schatzman1993,Zahn1997}.
        
    \subsection{Stochastic excitation of gravity waves}
        The stochastic excitation of gravity waves is a process that occurs in stars with a radiative-convective interface (both from a convective core and a convective envelope envelope). At this interface, turbulent motions can generate gravity waves that propagate into the radiative zone. An order of magnitude of the energy flux of these waves can be computed based on the local properties of the star at this interface as in \cite{Press1981} \citep[see also][]{Lecoanet2013}:
        \begin{equation}
            F_E^{S} = \rho_{int}v^3_{c, int} \frac{\omega_c}{N_{int}}\ ,
        \end{equation}
        where $S$ stands for stochastic, $\rho_{int}$ is the density and $v_{c, int}$ the convective velocity at the radiative-convective interface, $\omega_c$ the angular frequency of the wave (the convective frequency in this case) and $N_{int}$ the Brunt-Väisälä frequency at the location where the waves are excited. This flux can be integrated over the entire interface (in this case a spherical interface of radius $r_{int}$) to obtain the energy luminosity:
        \begin{equation}\begin{aligned}
            L_E^{S} &= \int_0^{2\pi}\int_0^\pi F_E^{S}(r, \theta, \varphi) r^2 \sin \theta \dd \theta \dd \varphi \\&= F_E^{S} \int_0^{2\pi}\int_0^\pi r^2 \sin \theta \dd \theta \dd \varphi \\&= 4 \pi r_{int}^2 \rho_{int}v^3_{c, int} \frac{\omega_c}{N_{int}}\ ,
        \end{aligned}\end{equation}
        where $\text{\Large $\sfrac{\omega_c}{N_{int}}$}$ represents the Froude number. As this quantity is not always trivial to compute, it has been approximated to be roughly equal to the Mach number at the radiative-convective interface \citep[see e.g.][]{Fuller2014}. For this reason the Froude number is sometimes called the convective Mach number in the literature. An other approach is to evaluate $N_{int}$ directly at the radiative-convective interface. A comparisson is performed in Fig. \ref{fig:Froude}, showing that both approaches lead to different results for the energy and angular momentum luminosities carried by stochastically-excited waves by up to two orders of magnitude. However, when comparing both approaches with those transported by tidally-excited waves remains robust because of their strong dependance on the orbital separation (see Fig. \ref{fig:M20L_cs}).

        To compute the Brunt-Väisälä frequency at the location where the waves are excited we can linearly expand the (squared) Brunt-Väisälä frequency starting from the interface where this frequency is zero. We thus only consider the first order term:
        \begin{equation}
            N_{int}^2 = 0 + \left.\frac{\dd N^2}{\dd r}\right|_{r_{int}} \lambda = \left.\frac{\dd N^2}{\dd\ln r}\right|_{r_{int}} \frac{\lambda}{r_{int}}
        \end{equation}
        where $\lambda$ is a characteristic length scale of the variation in the gravity wave in the radial direction, defined as in \cite{Goodman1998} and \cite{Ahuir2021a}:
        \begin{equation}
            \lambda = \omega_c^\frac{2}{3}{(l(l+1))}^{-\frac{1}{3}}\left|\frac{\dd N^2}{\dd \ln r}\right|^{-\frac{1}{3}}_{r_{int}}r_{int}\ .
        \end{equation}
        Therefore we obtain Brunt-Väisälä frequency at the location where the waves are excited as
        \begin{equation}
            N_{int} = \omega_c^\frac{1}{3}{(l(l+1))}^{-\frac{1}{6}}\left|\frac{\dd N^2}{\dd \ln r}\right|^\frac{1}{3}_{r_{int}}\ .
        \end{equation}
        This allows us to rewrite the energy luminosity
        \begin{equation}\begin{aligned}
            L_E^{S} &= 4 \pi r_{int}^2 \rho_{int}v^3_{c, int} \frac{\omega_c}{\omega_c^\frac{1}{3}{(l(l+1))}^{-\frac{1}{6}}\left|\frac{\dd N^2}{\dd \ln r}\right|^\frac{1}{3}_{r_{int}}}\\
            &= 4 \pi {(l(l+1))}^\frac{1}{6} \rho_{int} r_{int}^2 v^3_{c, int}\left|\frac{\dd N^2}{\dd \ln r}\right|^{-\frac{1}{3}}_{r_{int}}\omega_c^\frac{2}{3}\ .
        \end{aligned}\end{equation}
        In this case $\omega_c$ is the angular frequency of the convection, thus $\omega_c = 2\pi/t_c$ with $t_c = l_c/v_c$ the convective turnover time, computed at $r_{int}$, finally resulting in:
        \begin{equation}\label{eq:LES}
            L_E^{S} = 4 \pi {(l(l+1))}^\frac{1}{6} \rho_{int} r_{int}^2 v^3_{c, int}\left|\frac{\dd N^2}{\dd \ln r}\right|^{-\frac{1}{3}}_{r_{int}}\left(\frac{2\pi}{t_{c, int}}\right)^\frac{2}{3}\ .
        \end{equation}
        When calculating the angular momentum luminosity $L_L^{S}$, this becomes
        \begin{equation}\label{eq:LLS}\begin{aligned}
            L_L^{S} &= \frac{m}{\omega_c} L_E^{S} \\&= 4 \pi m {(l(l+1))}^\frac{1}{6} \rho_{int} r_{int}^2 v^3_{c, int}\left|\frac{\dd N^2}{\dd \ln r}\right|^{-\frac{1}{3}}_{r_{int}}\left(\frac{2\pi}{t_{c, int}}\right)^{-\frac{1}{3}}\ .
        \end{aligned}\end{equation}
        The absolute value of $L_L$ gives the global torque applied on the studied radiative zone as progressive gravity waves are damped before reaching its other boundary (we refer the reader to Appendix E in \citealp{Ahuir2021a}).

        In practice when computing these luminosities, the convective properties ($v_{c, int}$ and $t_{c, int}$) and the change in the Brunt-Väisälä frequency ($\dd N^2/\dd \ln r$) are evaluated at the radiative-convective interface where the waves are launched. Here the mean of the properties through the first 30 datapoints inside the convective zone (for $v_{c, int}$ and $t_{c, int}$) and inside the radiative zone (for $\dd N^2/\dd \ln r$) is used for more stable values for these properties.
        For interpretation purposes, it is useful to rewrite the convective velocity using the approximate formulation from mixing length theory $v_c^3 = L_\star / (\rho_\text{CZ} R_\star^2)$, with $L_\star$ the luminosity of the star, $\rho_\text{CZ}$ the mean density in the convective zone, and $R_\star$ the radius of the star \citep{Brun2017}.

        This expression for the angular momentum luminosity strongly relies on the monochromatic model from \cite{Press1981} where it is assumed that convection excites gravity waves efficiently for $\omega\approx\omega_c$. It has been later refined with taking into account the possibility to excite a frequency spectrum \citep{GarciaLopez1991, Zahn1997}, the contribution of the Reynolds stresses in the bulk of stellar convective regions \citep{Goldreich1990,Belkacem2009}, the nature of the stratification gradient at the radiative/convective interface \citep{Lecoanet2013}, and convective plumes \citep{Schatzman1993, Pincon2016}. The obtained predictions always scale with the vertical flux of kinetic energy carried by convection and the Froude number at the convective/radiative interface as proposed by \cite{Press1981}. At the same time, strong efforts have been done to develop and compute local and global hydrodynamical nonlinear simulations of internal gravity waves excitation by turbulent convection in stellar interiors \citep{Rogers2013,Alvan2014,Couston2018,Edelmann2019,LeSaux2022,Breton2022,Anders2023,Herwig2023,Daniel2025}. Although these simulations provide a more realistic description of the excitation process, they have confirmed that theoretical models, among which the model proposed by Press (1981), provide a reasonable order of magnitude estimate of the total wave flux \citep{Rogers2013,Alvan2014,Couston2018,Edelmann2019,Daniel2025}.
        
        Additionaly, rotation itself can have an impact on the stochastic excitation of gravity waves through the formation of gravito-inertial waves \citep[e.g.][]{Augustson2020}, or even magneto-gravito-inertial waves when magnetic fields are present \citep[e.g.][]{Mathis2009,Mathis2012,Bessila2024}. However, modelling (magneto-)gravito-inertial waves is out of scope for this work and we refer the reader to Sect. 4 for a detailed discussion on the required developments. We will rely on the simpler model from \cite{Press1981} in the remainder of this work as a first estimate.

    \subsection{Tidal excitation of gravity waves}
        Tidal forces from a companion can excite gravity waves at the radiative-convective interface that propagate into the radiative zone. The tidal perturbation can be described by the tidal potential \citep{Ogilvie2014}:
        \begin{equation}
            \begin{aligned}
                \Psi(r, \theta, \varphi, t)=\Re \Bigg[\sum_{l=2}^{\infty} \sum_{m=0}^l \sum_{n=-\infty}^{\infty}\Bigg\{\frac{G M_2}{a} & A_{l, m, n} (e, i)\left(\frac{r}{a}\right)^l\\& \times Y_l^m(\theta, \varphi) \mathrm{e}^{-\mathrm{i} n \Omega_o t}\Bigg\}\Bigg]\ ,
            \end{aligned}
        \end{equation}
        where $G$ is the gravitational constant, $M_2$ the mass of the companion, $a$ the semi-major axis of the orbit, $A_{l, m, n} (e, i)$ a coefficient depending on the eccentricity $e$ and inclination $i$ of the orbit relatively to the stellar spin, $Y_l^m(\theta, \varphi)$ the spherical harmonic function with degree $l$ and order $m$, $\Omega_o$ the orbital frequency, and $\omega_t = n\Omega_o - m\Omega_s$ the tidal frequency in the frame rotating with the star with $\Omega_s$ the rotation frequency of the star. In this work we will only consider circular ($e=0$) and coplanar ($i=0$) orbits, and thus only consider the dominant quadrupolar term ($l=m=2$) with $n=2$, resulting in the following tidal potential:
        \begin{equation}\label{eq:TidalPotential}
            \begin{aligned}
                \Psi=&\operatorname{Re} \left\{\frac{G M_2}{a} \sqrt{\frac{6\pi}{5}}\left(\frac{r}{a}\right)^2 Y_{2}^{2}(\theta, \varphi) \mathrm{e}^{-\mathrm{i} 2 \Omega_o t} \right\}\\
                 \equiv &\ \varphi_T(r)\operatorname{Re} \left\{ Y_{2}^{2}(\theta, \varphi) \mathrm{e}^{-\mathrm{i} 2 \Omega_o t} \right\}\ .
            \end{aligned}
        \end{equation}
        This tidal potential can excite gravity waves at the radiative-convective interface, which then propagate through the radiative zone. The energy and angular momentum luminosities of these waves has been computed by \cite{Ahuir2021a}:
        \begin{align}
            L_E^{T} &= \frac{3^\frac{2}{3}\Gamma^2\left(\frac{1}{3}\right)}{8\pi}\omega_t^{\frac{11}{3}}{(l(l+1))}^{-\frac{4}{3}}\rho_{int}r_{int} \left|\frac{\dd N^2}{\dd \ln r}\right|^{-\frac{1}{3}}_{r_{int}} \mathcal{F}^2\label{eq:LET}\\
            L_L^{T} &= \frac{3^\frac{2}{3}\Gamma^2\left(\frac{1}{3}\right)}{8\pi}\omega_t^{\frac{8}{3}}m{(l(l+1))}^{-\frac{4}{3}}\rho_{int}r_{int} \left|\frac{\dd N^2}{\dd \ln r}\right|^{-\frac{1}{3}}_{r_{int}} \mathcal{F}^2\label{eq:LLT}
        \end{align}
        where $\Gamma$ is the gamma function, and $\mathcal{F}$ is the tidal forcing defined for both a convective core and a convective envelope as (see Fig. \ref{fig:TriLayer} for a schematic representation of a three-layer stellar structure):
        \begin{figure}
            \centering
            \includegraphics[width=\linewidth]{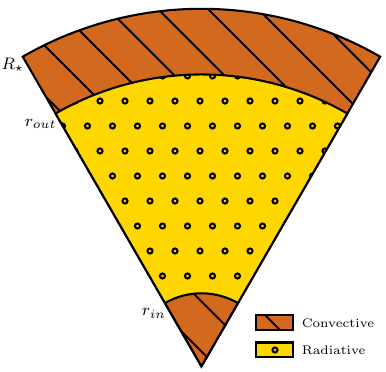}
            \caption{Schematic view of the radiative and convective shells in the three-layer model used in this work. Radii are not to scale (see Kippenhahn diagrams, e.g. Fig. \ref{fig:Kippenhahn}).}\label{fig:TriLayer}
        \end{figure}
        \begin{equation}\label{eq:TidalForcing}
            \begin{aligned}
            \mathcal{F}_\text{{in}}&=\int_0^{r_\text{in}}\left[{\left(\frac{r^2 \varphi_T}{g_0}\right)}^{\prime \prime}-\frac{l(l+1)}{r^2}\left(\frac{r^2 \varphi_T}{g_0}\right)\right] \frac{X_{1, \text {in}}}{X_{1, \text {in}}\left(r_\text{{in}}\right)} \dd r \\
            \mathcal{F}_\text{{out}}&=\int_{r_\text{{out}}}^{R_{\star}}\left[{\left(\frac{r^2 \varphi_T}{g_0}\right)}^{\prime \prime}-\frac{l(l+1)}{r^2}\left(\frac{r^2 \varphi_T}{g_0}\right)\right] \frac{X_{1, \text {out}}}{X_{1, \text {out}}\left(r_\text{{out}}\right)} \dd r
            \end{aligned}\ ,
        \end{equation}
        where $r_\text{in}$ and $r_\text{out}$ are the inner and outer boundaries of the radiative zone, respectively, and $X_{1, \text {out}}$ and $X_{1, \text {in}}$ are representations for the radial displacement originating from the inner and outer boundary of the radiative zone. The radial displacement can be calculated using the following differential equations and boundary conditions \citep{Ahuir2021a,Esseldeurs2024}:
        \begin{equation}
            \begin{aligned}
                &\left\{\begin{aligned}
                    & X_{1, \text {out}}^{\prime \prime}-\frac{\partial_r \rho_0}{\rho_0} X_{1, \text {out}}^{\prime}-\frac{l(l+1)}{r^2} X_{1, \text {out}}=0 \\
                    & X_{1, \text {out}}{(r)}_{r\to0} \propto r^{1/2+\sqrt{1/4+l(l+1)}} \\
                    & X_{1, \text {out}}^{\prime}{(r)}_{r\to0} \propto \left(1/2+\sqrt{1/4+l(l+1)}\right) r^{-1/2+\sqrt{1/4+l(l+1)}}
                \end{aligned}\right.\\
                &\left\{\begin{aligned}
                    & X_{1, \text {in}}^{\prime \prime}-\frac{\partial_r \rho_0}{\rho_0} X_{1, \text {in}}^{\prime}-\frac{l(l+1)}{r^2} X_{1, \text {in}}=0 \\
                    & X_{1, \text {out}}{(r)}_{r\to R_\star} \propto \rho_0\left(r-R_\star-\frac{\varphi_T(R_\star)}{g_0(R_\star)}\right) \\
                    & X_{1, \text {out}}^{\prime}{(r)}_{r\to R_\star} \propto \rho_0(R_\star)\ , 
                \end{aligned}\right.
            \end{aligned}
        \end{equation}
        where the proportionality factor in the boundary conditions is cancelled out as $X_j$ (with $j \in \{\text{in, out}\}$) is always rescaled to the interaction region $X_j/X_j(r_{int})$ in Eq.~\eqref{eq:TidalForcing}.

        Similar to stochastically excited gravity waves, tidally excited gravity waves can be influenced by rotation. This can lead to the formation of tidally excited (gravito-)inertial waves \citep[e.g.][]{Ogilvie2004,Ogilvie2007}, with a modification of the tidal frequency (as $\omega_t = n\Omega_o - m\Omega_s$). Modelling these effects will require bi-dimensional hydrodynamical numerical simulations \citep[e.g.][]{Ogilvie2004,Ogilvie2007,Dhouib2024}, which we discuss in detail as required developments in a near future in Sect. \ref{sec:discussion}, coupled with simulations of the secular rotational evolution of stars \citep[e.g.][for steps in this direction]{Mathis2015,Bolmont2016,Gallet2017,Ahuir2021b}, and we will rely on the non-rotating model from \cite{Ahuir2021a} in the remainder of this work as a first estimate.

    \subsection{Evaluating the competition}
        The competition between the stochastic and tidal excitation of gravity waves can be evaluated either by computing the luminosities for both processes, and by computing the ratio of the two relevant physical quantities. This ratio for the energy luminosities can be computed as
        \begin{equation}
            \frac{L_E^{T}}{L_E^{S}} = K \left(\frac{M_2}{M_1}\right)^2\frac{R_\star^{10}}{a^6} \omega_t^{\frac{11}{3}} \frac{1}{r_{int}v^{3}_{c, int}} {\left(\frac{2\pi}{t_{c, int}}\right)}^{-\frac{2}{3}}\tilde{\mathcal{F}}^2\ ,
        \end{equation}
        with $\mathcal{F} = \text{\Large$\sfrac{M_2}{M_1}$}\text{\Large$\sfrac{1}{a^3}$}\sqrt{\text{\Large$\sfrac{6 \pi}{5}$}}\ R_\star^5\tilde{\mathcal{F}}$ the unitless tidal forcing (see Appendix~\ref{app:TidalForcing}) and the constant $K$ defined as
        \begin{equation}
            K = \frac{3^\frac{5}{3}\Gamma^2\left(\frac{1}{3}\right)}{80\pi}{(l(l+1))}^{-\frac{3}{2}}\ ,
        \end{equation}
        and for the angular momentum luminosities as
        \begin{equation}\label{eq:ratio}
            \begin{aligned}
                \frac{L_L^{T}}{L_L^{S}} &= \frac{mL_E^{T}}{\omega_t}\frac{\omega_c}{mL_E^{S}}\\
                &= K \left(\frac{M_2}{M_1}\right)^2\frac{R_\star^{10}}{a^6} \omega_t^{\frac{8}{3}} \frac{1}{r_{int}v^{3}_{c, int}} {\left(\frac{2\pi}{t_{c, int}}\right)}^{\frac{1}{3}} \tilde{\mathcal{F}}^2\ .
            \end{aligned}
        \end{equation}
        This ratio therefore scales with the square of the mass ratio $q = M_2/M_1$, the tenth power of the stellar radius $R_\star$, and the inverse sixth power of the semi-major axis $a$, while it is independent of the stiffness in the Brunt-Väisälä frequency at the radiative-convective interface (stochastically-excited and tidally-excited gravity waves ``see'' the same stiffness of the Brunt-Väisälä frequency at the convection/radiation interface; so this independence can be expected).

    \subsection{Critical mass and critical orbital period}
        The critical mass of the companion is the mass beyond which the tidal excitation of gravity waves becomes stronger than the stochastic excitation. This can be computed by computing the companion mass for which the ratio of the angular momentum luminosities (Eq. \ref{eq:ratio}) is one. The critical mass therefore is computed as
        \begin{equation}
            M_\mathrm{crit} = \sqrt{K} M_1\frac{a^3}{R_\star^{5}} \omega_t^{-\frac{4}{3}} r_{int}^\frac{1}{2}v^{\frac{3}{2}}_{c, int} {\left(\frac{2\pi}{t_{c, int}}\right)}^{-\frac{1}{6}} \tilde{\mathcal{F}}^{-1}\ .
        \end{equation}
        Similarly, the critical orbital period can be computed by setting the two angular momentum luminosities equal to each other, and solving for the orbital period $P_\mathrm{orb} = 2\cdot2\pi/\omega_t$ (where $n=2$ is applied). This results in
        \begin{equation}
            P_\mathrm{orb, crit} = 4\pi \Bigg[K^{-\frac{3}{8}} \left(\frac{M_1}{M_2}\right)^{\frac{3}{4}} R_\star^{-\frac{15}{4}} a^{\frac{9}{4}} r_{int}^{-\frac{3}{8}} v^{\frac{9}{8}}_{c, int}{\left(\frac{2\pi}{t_{c, int}}\right)}^{-\frac{1}{8}} \tilde{\mathcal{F}}^{-\frac{3}{4}}\Bigg]\ .
        \end{equation}
        For $P_\mathrm{orb}>P_\mathrm{orb,crit}$, the angular momentum carried by gravity waves excited by stochastic excitation and the related torque is dominating the tidally excited ones and vice-versa.

\section{Competition between gravity waves excited by convection and tides}\label{sec:competition}
        Both the stochastic excitations and the tidal excitations depend on the internal structure of the star. Therefore, different initial masses for the host star will have different internal structures, and therefore the strengths of the excitations will be different. In this work we consider two different initial masses for the host star: 1 M$_\odot$ (late type stars; Sect.~\ref{sec:late-type}) and 2 M$_\odot$ (early type stars; Sect.~\ref{sec:early-type}), with the stellar evolutionary models described in \cite{Esseldeurs2024}\footnote{A small description can be found in Appendix~\ref{app:StellarEvolution}.}. Other initial masses (1.2, 1.4, 1.6, 1.8, 2.5, 3, 3.5 and 4 M$_\odot$) have also been treated, but as the results are similar to the 1 M$_\odot$ and 2 M$_\odot$ models, they are not shown here but can be found on \url{https://zenodo.org/records/19257979}.

    \subsection{Late type binary stars}\label{sec:late-type}
        \subsubsection{Internal structure}\label{sec:internal-structure}
            \begin{figure}
                \centering
                \includegraphics[width=\linewidth]{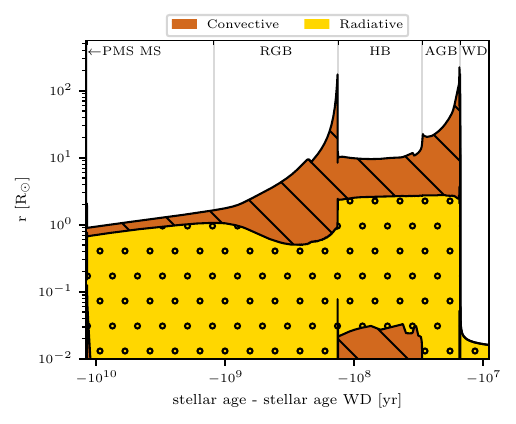}
                \caption{Kippenhahn diagram throughout the evolutionary stages of a $M_\mathrm{ZAMS} = 1$ M$_\odot$ star. All stellar evolutionary phases are indicated at the top. The yellow dotted region illustrates the radiative regions while the brown hatched region illustrates the convective regions. Stellar evolutionary phases (PMS to WD) are indicated.}\label{fig:Kippenhahn}
            \end{figure}
            
            The Kippenhahn diagram during the entire evolution of a $M_\mathrm{ZAMS} = 1$ M$_\odot$ late type star can be found in Fig.~\ref{fig:Kippenhahn}. The $x$-axis is shown in stellar age up to the stellar age where the star is a faint white dwarf (WD) with a luminosity of $L_\star = 10^{-1}$ L$_\odot$. During the main-sequence (MS) phase, the star has a convective envelope and a radiative core. When the hydrogen in the core is depleted, hydrogen shell burning starts and the star becomes a red giant branch (RGB) star. In this phase, the convective envelope expands, increasing the radius of the star. During this process the temperature in the core increases, eventually igniting helium burning in the core during the helium flash. After the helium flash the star becomes a horizontal branch (HB) star, with a three-layer structure (convective core, radiative envelope, and convective envelope). After the HB phase the star becomes an asymptotic giant branch (AGB) star, with again a deep convective envelope and a radiative core. The star loses mass during the RGB and AGB phases, and eventually becomes a WD.
        
        \subsubsection{Angular momentum luminosities}
            \begin{figure}
                \centering
                \includegraphics[width=\linewidth]{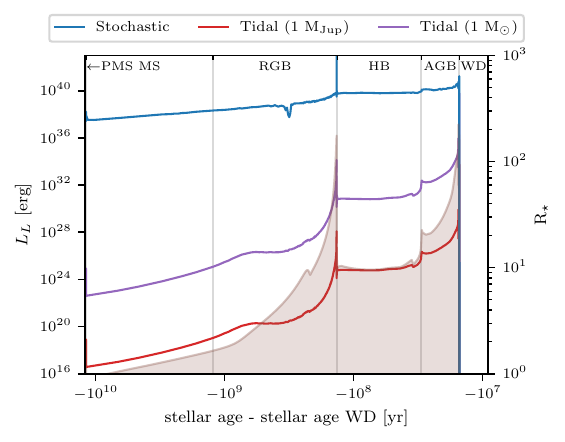}
                \caption{Angular momentum luminosities carried by stochastically (blue) and tidally (red and purple) excited gravity waves (left axis) and stellar radius (brown; right axis) as a function of stellar age for a $M_\mathrm{ZAMS} = 1$ M$_\odot$ star with a 1 M$_\mathrm{Jup}$ (red) and a 1 M$_\odot$ (purple) companion orbiting at 1 AU. Stellar evolutionary phases (PMS to WD) are indicated.}\label{fig:stoch}
            \end{figure}
            \begin{figure}
                \centering
                \includegraphics[width=\linewidth]{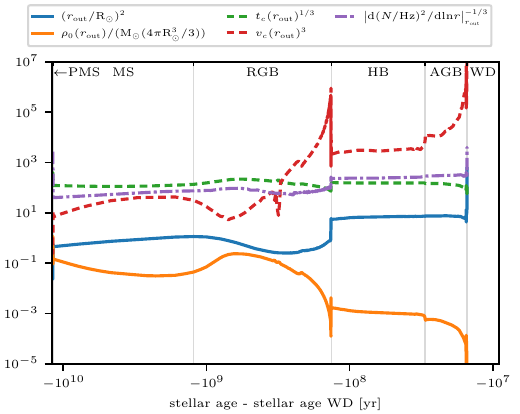}
                \caption{Different parameters influencing the angular momentum luminosity carried by stochastically excited gravity waves (Eq.~\ref{eq:LLS}) as a function of stellar age for a $M_\mathrm{ZAMS} = 1$ M$_\odot$ star. The change in radius of the radiative-convective boundary, density, convective timescale, convective velocity, and change in Brunt-Väisälä frequency and Brunt-Väisälä frequency squared at this radius are represented in blue (solid) orange (solid), green (dashed), red (dashed), purple (dash-dotted) and brown (dash-dotted), respectively. Stellar evolutionary phases (PMS to WD) are indicated.}\label{fig:compStoch}
            \end{figure}
            \begin{figure}
                \centering
                \includegraphics[width=\linewidth]{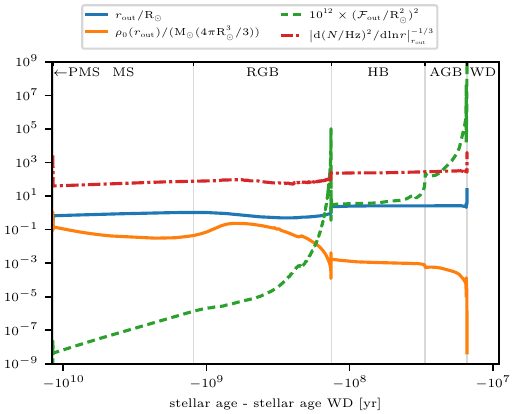}
                \caption{Different parameters influencing the angular momentum luminosity carried by tidally excited gravity waves (Eq.~\ref{eq:LLT}) as a function of stellar age for a $M_\mathrm{ZAMS} = 1$ M$_\odot$ star with a 1 M$_\mathrm{Jup}$ companion orbiting at 1 AU. The change in radius of the radiative-convective boundary is represented in blue (solid), the change in  density at this radius is represented in orange (solid), the change in tidal forcing is represented in green (dashed) and the change in Brunt-Väisälä frequency squared is represented in red (dash-dotted). Stellar evolutionary phases (PMS to WD) are indicated.}\label{fig:compTides}
            \end{figure}

            The angular momentum luminosities carried by stochastically excited and tidally excited gravity waves are shown in Fig.~\ref{fig:stoch} for a Jupiter mass companion with an orbital period of 1 year (used for ilustration purposes, a companion further than the maximal radius of the host star). The stochastic excitation dominates the tidal excitation during the entire evolution of the star, but this is strongly dependent on the orbital period, which will be investigated further in later sections.
            
            Overall the angular momentum luminosity carried by stochastically excited gravity waves is rather constant (compared to the angular momentum luminosity carried by tidally excited gravity waves), increasing from $10^{37}$ erg during the MS to $10^{40}$ erg when the star enters later evolutionary phases. During the entire evolution the luminosity is dominated by the waves excited from the convective envelope to the radiative core, except during the helium flash where the intense luminous flash increases the convective velocity in the convective core by several orders of magnitude. This results in a short spike in the convective velocity at the convective-radiative interface of the core. This spike results in a short burst in the angular momentum luminosity, although this burst does not last long enough to contribute to the overall transport. To understand the influence of the other physical parameters, the evolution of the relevant parameters for the angular momentum luminosity (see Eq.~\ref{eq:LLS}) can be seen in Fig.~\ref{fig:compStoch}. Here, mostly the convective velocity (in red dashed) and the density at the interface (in orange solid) are dominating changes in the angular momentum luminosity. They are anti-correlated, which can be understood by looking at the definition of the convective velocity in the standard mixing-length theory as $v_c^3 = L_\star / (\rho_\text{CZ} R_\star^2)$. The remaining dominant factor in the convective velocity is the stellar luminosity, which increases during the later evolutionary phases. Therefore the convective velocity will also gradually increase throughout the later evolutionary phases compared to the local density at the interface, and thus the angular momentum luminosity carried by stocchastically-excited gravity waves will increase as well.

            The angular momentum luminosity carried by tidally-excited gravity waves is much lower than the stochastic excitation (in the case of $P_\textrm{orb} = 1$ year), but increases more drastically during the later evolutionary phases. The evolution of the relevant parameters influencing this angular momentum luminosity (see Eq.~\ref{eq:LLT}) can be seen in Fig.~\ref{fig:compTides}. Here it can be seen that the tidal forcing is the dominant factor increasing by multiple orders of magnitude throughout the evolved phases. The strongest component influencing the tidal forcing evolution is the evolution of the stellar radius, because the tidal forcing ($\mathcal{F}$) is dependent on $R_\star^5$ \citep[see][and Appendix~\ref{app:TidalForcing}]{Esseldeurs2024}. This is counteracted a bit by the density at the interface, which decreases during the evolved phases, but not enough to counteract the increase in the tidal forcing.

        \subsubsection{The case of star-planet systems}\label{sec:star-planet-late}
            \begin{figure}
                \centering
                \includegraphics[width=\linewidth]{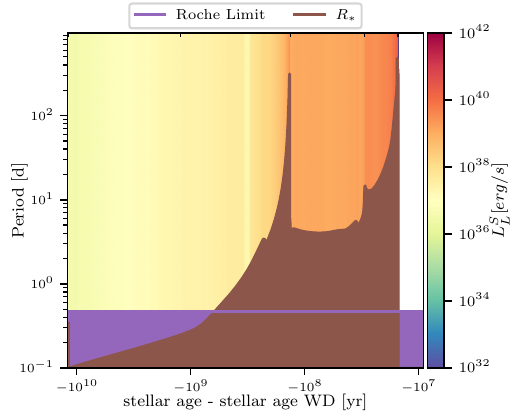}
                \caption{Angular momentum luminosity carried by stochastically excited gravity waves (Eq.~\ref{eq:LLS}) as a function of stellar age for a $M_\mathrm{ZAMS} = 1$ M$_\odot$ star and orbital period of a 1 M$_\mathrm{Jup}$ companion. $R_\star$ (in brown) represents the orbital period on which a companion orbits at the surface of the star, and the Roche limit is given in purple. Changes in the stellar evolutionary phase are indicated with ticks on the upper axis.}\label{fig:LLS}
            \end{figure}
            \begin{figure}
                \centering
                \includegraphics[width=\linewidth]{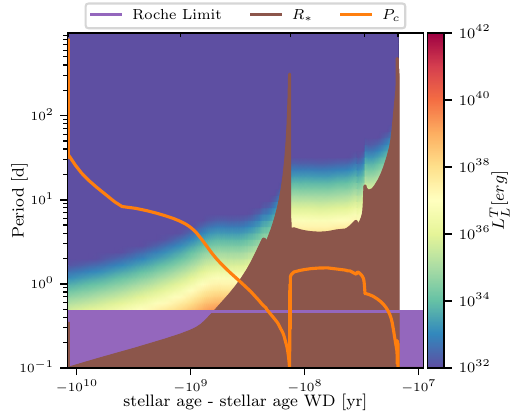}
                \caption{Angular momentum luminosity carried by tidally excited gravity waves (Eq.~\ref{eq:LLT}) as a function of stellar age for a $M_\mathrm{ZAMS} = 1$ M$_\odot$ star and orbital period of a 1 M$_\mathrm{Jup}$ companion. The critical period, $P_c$, above which companions excite progressive IGWs, is shown in orange. $R_\star$ (in brown) represents the orbital period on which a companion orbits at the surface of the star, and the Roche limit is given in purple. Changes in the stellar evolutionary phase are indicated with ticks on the upper axis.}\label{fig:LLT}
            \end{figure}
            \begin{figure}
                \centering
                \includegraphics[width=\linewidth]{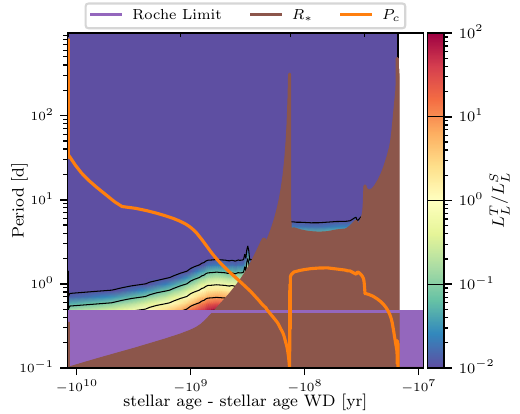}
                \caption{Ratio of the angular momentum luminosity carried by tidally excited gravity waves (Eq.~\ref{eq:LLT}) to the angular momentum luminosity carried by stochastically excited gravity waves (Eq.~\ref{eq:LLS}) as a function of stellar age for a $M_\mathrm{ZAMS} = 1$ M$_\odot$ star with a Jupiter mass companion. Contour lines are shown for ratios of 0.01, 0.1, 1, 10 and 100. The middle contour line with ratio 1 indicates the critical orbital period $P_\mathrm{orb, crit}$ for this companion mass. The critical period, $P_c$, above which companions excite progressive IGWs, is shown in orange. $R_\star$ (in brown) represents the orbital period on which a companion orbits at the surface of the star, and the Roche limit is given in purple. Changes in the stellar evolutionary phase are indicated with ticks on the upper axis.}\label{fig:Jup}
            \end{figure}

            For an orbital period of 1 year (used for ilustration purposes, a companion further than the maximal radius of the host star), a Jupiter mass planet will not have a significant impact on the angular momentum luminosities inside the star. This, however, strongly depends on the orbital period of the object. This can be seen in Fig.~\ref{fig:LLT}, where the angular momentum luminosity carried by tidally-excited gravity waves is shown as a function of stellar age and orbital period. Here the Roche limit (in purple) as well as the critical period (in orange) are plotted. The Roche limit represents the radius (or orbital period) at which the tidal forces on the planet are sufficiently strong to disrupt the planet (for the calculation, we refer to \citealp{Benbakoura2019}). The angular momentum depends on the tidal frequency (and thus the orbital frequency) to the power of 8/3 (see Eq. \ref{eq:LLT}). Therefore when the orbital period decreases, the orbital frequency and thus the tidal frequency increases, increasing the tidal angular momentum luminosity. The ratio of the angular momentum luminosity carried by tidally-excited gravity waves to the angular momentum luminosity carried by stochastically-excited gravity waves is shown in Fig.~\ref{fig:Jup}. Here it can be seen that for a Jupiter mass companion the stochastic excitations completely dominate over tidal excitations in most of the parameter space. However for orbital periods shorter than about a day at the beginning of the RGB phase, the tidal excitations starts to have an influence. Although this is only a small part of the parameter space we explored, planets have been observed close to this regime \citep[e.g.][]{Jones2014,Saunders2024}.

        \subsubsection{The case of binary stars}\label{sec:star-star-late}
            \begin{figure}
                \centering
                \includegraphics[width=\linewidth]{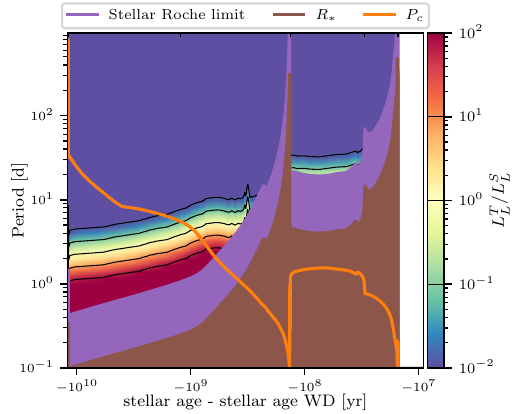}
                \caption{Ratio of the angular momentum luminosity carried by tidally excited gravity waves (Eq.~\ref{eq:LLT}) to the angular momentum luminosity carried by stochastically excited gravity waves (Eq.~\ref{eq:LLS}) as a function of orbital period for a solar mass companion.  Contour lines are shown for ratios of 0.01, 0.1, 1, 10 and 100. The middle contour line with ratio 1 indicates the critical orbital period $P_\mathrm{orb, crit}$ for this companion mass. The critical period, $P_c$, above which companions excite progressive IGWs, is shown in orange. $R_\star$ (in brown) represents the orbital period on which a companion orbits at the surface of the star, and the stellar Roche limit is given in purple. Changes in the stellar evolutionary phase are indicated with ticks on the upper axis.
                }\label{fig:Sun}
            \end{figure}

            When the mass of the companion increases, so will the angular momentum luminosity of the tidal excitation. This can be seen in Fig.~\ref{fig:Sun}, where ratio of the angular momentum luminosity carried by tidally excited gravity waves to the angular momentum luminosity carried by stochastically excited gravity waves is shown as a function of orbital period for a solar mass companion. Here the stellar Roche limit (distance at which the primary overfills its Roche lobe) is plotted rather than the Roche limit (distance at which the companion overfills its Roche lobe). In the figure it can be seen that the tidal excitation starts to contribute earlier, during the MS with orbital periods of a few days, and during the RGB with a period of just under 10 days. During the HB phase, the tidal excitations start to contribute for orbital periods of just under 30 days (a contribution larger than 1 \%), but to have an equal ratio the companion would have to be inside the stellar Roche limit.
        
        \subsubsection{Critical mass}\label{sec:Mcrit-late}
            \begin{figure}
                \centering
                \includegraphics[width=\linewidth]{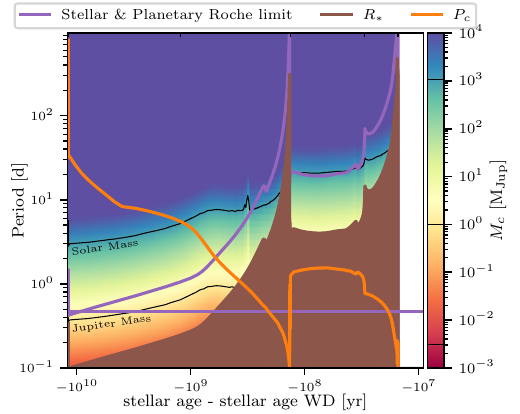}
                \caption{Critical mass as a function of the orbital period and stellar age for a $M_\mathrm{ZAMS} = 1$ M$_\odot$ star. The critical mass is the mass of the companion at which the angular momentum luminosity carried by tidally excited gravity waves (Eq.~\ref{eq:LLT}) is equal to the angular momentum luminosity carried by stochastically excited gravity waves (Eq.~\ref{eq:LLS}). Contour lines are shown for M$_\mathrm{Earth}$, M$_\mathrm{Jupiter}$ and M$_\odot$ companions. These contour lines also indicate thier critical orbital period. The critical period, $P_c$, above which companions excite progressive IGWs, is shown in orange. $R_\star$ (in brown) represents the orbital period on which a companion orbits at the surface of the star, and the stellar and planetary Roche limit is given in purple. Contour lines are shown for 1 M$_\mathrm{Earth}$, 1 M$_\mathrm{Jupiter}$ and 1 M$_\odot$. Changes in the stellar evolutionary phase are indicated with ticks on the upper axis.}\label{fig:M}
            \end{figure}

            This information about planetary and stellar companions can be combined using the critical mass. The critical mass as a function of orbital period and stellar age is shown in Fig.~\ref{fig:M}. Here both the stellar and planetary Roche limit are plotted. Contour lines are also shown for one Earth mass, one Jupiter mass, and one solar mass. The contour for the Earth mass does not show up in the diagram, indicating that an earth mass will never be able to excite gravity waves with an angular momentum luminosity larger than the already stochastically excited gravity waves. The Jupiter mass contour rises above its Roche limit at the beginning of the RGB phase (consistent with Sect.~\ref{sec:star-planet-late}), and the stellar mass contour ranges from 2 to 5 days in the MS and the beginning of the RGB, but during the HB it is not able to reach above its Roche limit (consistent with Sect.~\ref{sec:star-star-late}).

    \subsection{Early type binary stars}\label{sec:early-type}
        \begin{figure*}
            \centering
            \includegraphics[width=0.95\linewidth]{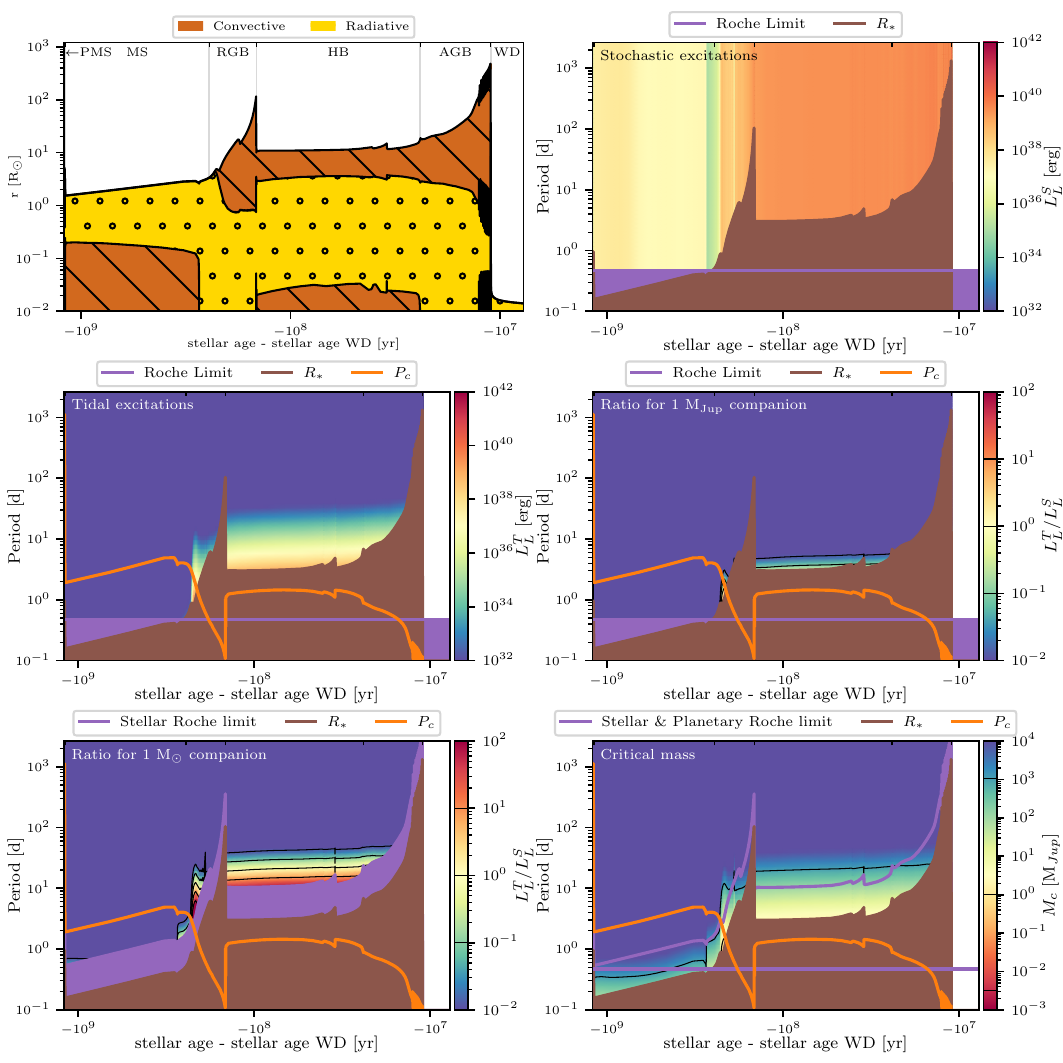}
            \caption{Internal structure and angular momentum luminosities for a $M_\text{ZAMS} = 2$ M$_\odot$ star. Top left: Kippenhahn diagram. The brown hatched regions represent convective layers, and the yellow dotted region represents radiative layers. Stellar evolutionary phases (PMS to WD) are indicated. Top right and center left: angular momentum luminosities carried by stochastically and tidally excited gravity waves as a function of orbital period and stellar age. Center right and bottom left: ratio of the angular momentum luminosity carried by tidally excited waves to the angular momentum luminosity carried by stochastically excited waves as a function of orbital period and stellar age for a 1 M$_\mathrm{Jup}$ (center right) and a 1 M$_\odot$ (bottom left) companion. Bottom right: critical mass as a function of orbital period and stellar age. The Roche limit is shown in purple, the critical period ($4\pi/\omega_c$; see Eq.~\ref{eq:CriticalFrequency}) in orange, and the period at which a planet orbits at the stellar radius in brown. Changes in the stellar evolutionary phase are indicated with ticks on the upper axis.}\label{fig:M20L}
        \end{figure*}
        \subsubsection{Internal structure}
            For stars with higher initial masses, the internal structure is different. The Kippenhahn diagram for a $M_\mathrm{ZAMS} = 2$ M$_\odot$ star can be seen in Fig.~\ref{fig:M20L} (top left). The star has a convective core and a radiative envelope during the MS phase. When the hydrogen in the core is depleted, the star turns into an RGB star, similar to the 1 M$_\odot$ star (see Sect.~\ref{sec:internal-structure}) following the same later evolution. The star has a helium flash at the end of the RGB turning it into a HB star, and then into an AGB star. The star loses mass during the RGB and AGB phases, and eventually becomes a WD.

        \subsubsection{Angular momentum luminosities}
            The angular momentum luminosities carried by stochastically and tidally excited gravity waves are shown in Fig.~\ref{fig:M20L} (top right and middle left panel respectively). Similar to the 1 M$_\odot$ star, the stochastic excitation remains (compared to the angular momentum luminosity carried by tidally excited gravity waves), with a slight increase from $10^{39}$ erg during the MS to $10^{40}$ erg when the star enters later evolutionary phases. Here however, there is a short time during the transition from MS to RGB where the angular momentum luminosity drops by 4 orders of magnitude. During the MS the angular momentum luminosity is dominated by the contribution from the convective core, as the star has a negligible convective envelope. At the end of the MS this convective core disappears, while the convective envelope only appears during the RGB phase itself. Therefore there is a transition period where there only is a negligible convective envelope, and no convective core.

            The angular momentum luminosity carried by tidally excited gravity waves during the MS is negligible compared to the angular momentum luminosity carried by stochastic excitation. During the RGB when the convective envelope appears, the angular momentum luminosity carried by tidally excited gravity waves increases, but remains relatively small during its entire evolution.

        \subsubsection{The case of star-planet systems}\label{sec:star-planet-early}
            For a Jupiter mass companion, the angular momentum carried by tidally-excited gravity waves is negligible compared to the angular momentum carried by stochastically-excited waves during the entire evolution. This can be seen in Fig.~\ref{fig:M20L} (middle left panel), where the ratio of the angular momentum luminosity carried by tidally excited gravity waves to the angular momentum luminosity carried by stochastically excited gravity waves is shown as a function of stellar age.

        \subsubsection{The case of binary stars}\label{sec:star-star-early}
            For a solar mass companion, the angular momentum carried by tidally-excited gravity waves is negligible compared to the angular momentum carried by stochastically-excited waves during most of the evolution. This can be seen in Fig.~\ref{fig:M20L} (bottom left panel) where the ratio of the angular momentum luminosity carried by tidally excited gravity waves to the angular momentum luminosity carried by stochastically excited gravity waves is shown as a function of orbital period for a solar mass companion. Here, however, at the beginning of the RGB, as well as close to the stellar Roche limit during the HB phase, the tidal excitations starts to compete. This region of the parameter space however is so close to the host star that these systems are not long-lived, and therefore not relevant for the evolution of the star.

            A few observed early type stars in close binaries have measured internal rotation rates from asteroseismology \citep[e.g.][]{Beck2018,VanReeth2022,Michielsen2023}. In particular KIC 4930889, studied by \cite{Michielsen2023}, is a 4 M$_\odot$ star with a 3 M$_\odot$ companion in a 18 day orbit. In their study they investigated the necessity of additional convective boundary mixing and angular momentum transport on top of the transport by stochastically excited gravity waves where they found a preference for either an exponentially decaying mixing profile in the near-core region or absence of additional near-core mixing. This is consistent with our findings that the tidal excitation of gravity waves is negligible compared to the stochastic excitation for this system.

\section{Discussion}\label{sec:discussion}
    \subsection{Transport of angular momentum}
        Given the results from Sect.~\ref{sec:competition}, it is clear that the stochastic excitation of gravity waves dominates over their tidal excitation for most of the parameter space. Only for very close-in stellar companions around late type stars does the tidal excitation start to compete with the stochastic excitation. For early type stars, the tidal excitation of gravity waves is negligible during the entire evolution, when compared to the stochastic excitation.

        This result has key implications. First of all, the presence of a companion will not significantly increase the angular momentum transport by gravity waves in the host star on evolutionary timescales, except for very close-in stellar companions. Once excited by a source term, an internal gravity wave will transport and redistribute angular momentum in the same way independently of its excitation mechanism (we refer the reader to \citealp{Talon2005} for stochastically-excited waves and to \citealp{Goldreich1989} and \citealp{Talon1998} for tidally-excited waves). Depending on the gradient of angular velocity between the place where they are excited and the propagation region we consider, and their prograde or retrograde behavior, they will deposit or extract angular momentum in the considered stellar radiation zone (for instance, a retrograde wave propagating in the radiative core of a solar-type star with a core rotating faster than the envelope will extract angular momentum). In this framework, the torques applied on the surrounding convective envelope by stellar winds, the equilibrium tide and tidal inertial waves in the case of late-type stars are essential since they are the source of the rotation gradient that leads to a net deposit or extraction of angular momentum from the radiative core. As showed by \cite{Goldreich1989} and \citep{Ahuir2021a}, the net global torque applied on the radiative region is then directly related to the wave angular momentum luminosity at the place of its excitation. Therefore, the redistribution of angular momentum by internal waves is a function of their angular momentum luminosity at the radiative/convective interface where they are excited (which is dominated by the stochastic excitation in a broad parameter space) and of the gradient between the rotation of the place we study in a stellar radiation zone and the rotation of the surrounding stellar convective envelope or surface, which results from the action of stellar winds and the equilibrium tide and tidal inertial waves in the case of a convective envelope \citep{Ahuir2021b}. If the torque applied by stellar winds dominates tidal torques while the stochastic excitation dominate the tidal one, the rotational evolution of a star will be weakly affected by the presence of the companion. This means that the presence of a companion will not help to explain the observed rotation rates of stellar cores in evolved stars, and other mechanisms are still needed to explain these observations. Second, as planetary companions do not significantly affect the angular momentum transport in the cores of their host stars, it is still possible to approximate the core rotational evolution of the host star without taking into account the presence of the planet\footnote{The envelope rotational evolution can still be effected by a planetary companion, see e.g. \cite{Benbakoura2019,Ahuir2021b}.}. This is especially useful when comparing asteroseismological observations to stellar evolution models, as the impact of a planet can be neglected for the internal wave-driven transport of angular momentum. Finally, as the tidal excitation of gravity waves is negligible in comparison with their stochastic excitation for most of the parameter space, it is not necessary to include this mechanism systematically in stellar evolution codes that already include the stochastic excitation of gravity waves \citep{Charbonnel2013,Mathis2013,Fuller2014}. This simplifies the implementation of angular momentum transport by gravity waves in these codes.
        However, the dissipation of tides will still be crucial because it is the sole vector that conveys global exchanges of angular momentum between the spin of the star and the orbit of its companion.
    
    \subsection{Impact of the Coriolis acceleration}
        The rotation of the star impacts the excitation, the propagation and the damping of both stochastically-excited and tidally-excited waves \citep[e.g.][]{Ogilvie2004,Ogilvie2007, Mathis2009, Mathis2014}.
        First, in convective regions, for sub-inertial frequencies (frequencies below the inertial frequency $2\Omega$), evanescent gravity waves found in the non-rotating case are replaced by propagative inertial waves that will be both excited by tides \citep[e.g.][]{Ogilvie2004,Ogilvie2007} and by turbulent convective Reynolds stresses \citep[e.g.][]{Bekki2022,Philidet2023,Blume2024,Fuentes2026}. The properties of inertial modes propagating in a spherical shell such as stellar convective envelopes require multi-D numerical simulations since the problem is bi-dimensional and non-separable \citep{Rieutord1997, Bekki2022}. Such heavy multi-D simulations are also required to compute their excitation by tides and by turbulent convection \citep{Ogilvie2004,Ogilvie2007, Bekki2022, Blume2024, Fuentes2026}. Analytical treatments of the tidal excitation and stochastic excitation are possible but at the price of strong assumptions such as an average over all possible tidal frequencies (and periods; \citealp{Ogilvie2013}) for the tidal excitation or of complex formalisms for the stochastic excitation \citep{Augustson2020,Neiner2020, Philidet2023}. In the case of super-inertial frequencies, the complexity also increases with gravito-inertial waves which are evanescent as gravity waves in the non-rotating case but with an evanescence that now depends on rotation \citep[e.g.][]{Mathis2014}. As a consequence, it is not yet possible to compare in a tractable and robust way the amplitude of tidally-excited and stochastically-excited waves propagating in stellar convective regions for the broad space of stellar and orbital parameters explored here.
        The same is true in stellar radiation zones where gravity waves become gravito-inertial waves with more complex propagation and damping as a function of the stratification and rotation \citep[e.g.][]{Lee1997, Mathis2009, Mirouh2016, Mathis2026}. Their tidal excitation becomes also more complex because of their couplings with tidal inertial waves at radiation/convection boundaries \citep{Ogilvie2004, Dhouib2024} that require the same multi-D numerical treatment that in stellar convective envelopes. Concerning the treatment of the stochastic excitation of gravito-inertial waves by turbulent motions in the adjacent convective regions, progresses have been achieved \citep{Mathis2014,Augustson2020,Neiner2020}. However, the variation of its strength as a function of rotation is complex. For instance, it decreases with increasing rotation for the radiative/convective interface excitation while it can increase for their excitation by Reynolds stresses in the bulk of convective regions \citep{Augustson2020}. These results have been obtained within a monochromatic approach, where the wave period is close to the characteristic convective turn-over time, and it must be extended to understand the possible impact of the turbulent spectrum.
        Taking into account the impact of rotation on the excitation, propagation and damping of both stochastically-excited and tidally-excited waves is therefore a complex task that requires multi-D numerical simulations and/or complex analytical treatments. This is out of the scope of this paper, but will be investigated in future work.

\section{Conclusion}\label{sec:conclusion}
    In this work, we have computed the angular momentum luminosities carried by stochastically and tidally excited gravity waves in stars with different initial masses. The angular momentum luminosity carried by stochastically excited gravity waves is much larger than the one carried by tidally excited waves, but this strongly depends on the orbital period of the companion. For a 1 M$_\odot$ star with a Jupiter mass companion, the tidal excitation of gravity waves only starts to compete with the stochastic excitation during the beginning of the RGB phase for orbital periods shorter than a day. For a solar mass companion, the tidal excitation of gravity waves only starts to compete with the stochastic excitation during the MS and RGB phases for orbital periods shorter than a few days. The critical mass of the companion is also computed, showing the same trend. For a 2 M$_\odot$ star, the tidal excitation of gravity waves is negligible when compared to the stochastic one during the entire evolution.

    Overall, our results suggest that while tidally excited IGWs are an interesting theoretical phenomenon, that are important for tidal interactions and the related orbital evolution, they do not play a major role in shaping the internal angular momentum evolution of low- and intermediate-mass stars with companions. Instead, stochastic excitation by convection remains the dominant IGW-driving mechanism. This simplifies forthcoming stellar modelling efforts, as the presence of companions can be neglected when studying IGW-driven angular momentum transport in stars.

    Note, however, that it still remains necessary to take into account that tidal torques applied to the convective envelope due to the equilibrium tide \citep{Zahn1966a,Zahn1989a,Remus2012} and to tidal inertial waves \citep{Ogilvie2004,Mathis2015,Barker2020}, which may compete with the torque applied by stellar winds \citep{Barker2009,Madappatt2016,Ahuir2021b}. In addition, this picture should be refined by considering the impact of both rotation and magnetic fields on the excitation, propagation and damping of stachastically and tidally excited magneto-gravito-inertial waves \citep[e.g.][]{Ogilvie2004,Ogilvie2007,Mathis2009,Mathis2012,Lin2018,Augustson2020,Bessila2024}.

\begin{acknowledgements}
    The authors would  like to thank the anonymous referee for their constructive comments which helped to improve the quality of the paper. M. Esseldeurs, S. Mathis and L. Decin acknowledge support from the FWO grant G0B3823N. M. Esseldeurs and L. Decin acknowledge support from the FWO grant G099720N, the KU Leuven C1 excellence grant MAESTRO C16/17/007, the KU Leuven IDN grant ESCHER IDN/19/028 and the KU Leuven methusalem SOUL grant METH/24/012. S. Mathis acknowledges support from the PLATO CNES grant at CEA/DAp, from the Programme National de Planétologie (PNP-CNRS/INSU) and from the European Research Council through HORIZON ERC SyG Grant 4D-STAR 101071505. L. Decin acknowledges support from the FWO sabbatical grant K803625N. While partially funded by the European Union, views and opinions expressed are however those of the author only and do not necessarily reflect those of the European Union or the European Research Council. Neither the European Union nor the granting authority can be held responsible for them.
\end{acknowledgements}

\bibliographystyle{aa}
\bibliography{bibliography}

\appendix
\section{Stellar evolution models}\label{app:StellarEvolution}
    In order to calculate the energy and angular momentum luminosities of gravity waves, we need to know the internal structure of a star throughout its lifetime. For this we used the stellar evolutionary code Modules for Experiments in Stellar Astrophysics \citep[MESA;][]{Paxton2011, Paxton2013, Paxton2015, Paxton2018, Paxton2019, Jermyn2023} using the same parameters as was done in \cite{Esseldeurs2024}\footnote{The inlist used to compute the stellar evolutionary models can be found both in \citealp{Esseldeurs2024} and on Zenodo: \url{https://doi.org/10.5281/zenodo.11519523}} for stars at solar metallicity ($Z = 0.0134$; \citealp{Asplund2009}) with initial masses between 1 and 4 M$_\odot$.

    Starting from the pre-main sequence (PMS), the stellar evolutionary models were computed up to the white dwarf (WD) stage where they are terminated when the luminosity reaches $L = 10^{-1} $ L$_\odot$. Convection in the simulations is modelled using the mixing length theory (MLT) following the prescription of \cite{Henyey1965} with $\alpha_{\mathrm{MLT}} = 1.931$ \citep{Cinquegrana2022} and opacity tables dedicated for low-temperature molecular opacities necessary in the evolved phases of evolution \citep[\AE{}SOPUS;][]{Marigo2009}. The atmosphere is simulated using a grey temperature-opacity relation based on the Eddington relation \citep{Paxton2011}. The mixing processes in the simulations are simplified dramatically to reduce the complexity of the models, and reduce the computational cost. In the radiative zones of the star a constant mixing coefficient $D_\text{min}=10$ cm$^2$ s$^{-1}$ is assumed for numerical stability.

    During the evolved phases, mass loss is taken into account. For the red giant branch (RGB) phase the Reimers prescription \citep{Reimers1975} is used with a scaling factor of $\eta_\text{Reimers}=0.477$ \citep{McDonald2015}. Further in the evolution during the asymptotic giant branch (AGB) phase the Bl{\"o}cker prescription \citep{Blocker1995} is used with a scaling factor of $\eta_\text{Bl{\"o}cker}=0.1$ for masses above 2 $M_\odot$ and $\eta_\text{Bl{\"o}cker}=0.05$ for masses below 2 $M_\odot$ \citep{Madappatt2016}.

\section{Unitless tidal forcing}\label{app:TidalForcing}
    The tidal forcing (Eq.~\ref{eq:TidalForcing}) is not a unitless quantity, as there are units in the tidal potential $\varphi_T$, the local gravity $g_0$, and the integration over the radius $r$. To understand the scaling of the tidal forcing, we can rewrite it to separate the unitless part from the part with units. The tidal forcing can be written as:
    \begin{equation}
        \mathcal{F} = \frac{M_2}{M_1}\frac{1}{a^3}\sqrt{\frac{6 \pi}{5}}R_\star^5\tilde{\mathcal{F}}\ ,
    \end{equation}
    where $\tilde{\mathcal{F}}$ is the unitless part of the tidal forcing given by
    \begin{align}
        \tilde{\mathcal{F}}_\text{out} &= \int_0^{\alpha}\left[\left(\frac{x^6}{m_r}\right)^{\prime \prime}-\frac{l(l+1)}{x^2}\left(\frac{x^6}{m_r}\right)\right] \frac{X_{1, \text {out}}}{X_{1, \text {out}}\left(\alpha\right)} \dd x\ ,\\
        \tilde{\mathcal{F}}_\text{in} &= \int_{\alpha_c}^1\left[\left(\frac{x^6}{m_r}\right)^{\prime \prime}-\frac{l(l+1)}{x^2}\left(\frac{x^6}{m_r}\right)\right] \frac{X_{1, \text {in}}}{X_{1, \text {in}}\left(\alpha_c\right)} \dd x\ ,
    \end{align}
    where $x = r/R_\star$ (the power 6 originates from 2 that is already present in the tidal forcing, 2 from the tidal potential and 2 from the local gravity), $\alpha = r_\text{out}/R_\star$ the envelope radius aspect ratio, $\alpha_c = r_\text{in}/R_\star$ the core radius aspect ratio, and $m_r = M_r / M_1$ with $M_r$ the mass coordinate as a function of radius. The tidal forcing therefore scales with the mass ratio of the companion to the star, the inverse of the semi-major axis cubed, and the radius of the star to the fifth power.

\begin{figure}
        \centering
        \includegraphics[width=\linewidth]{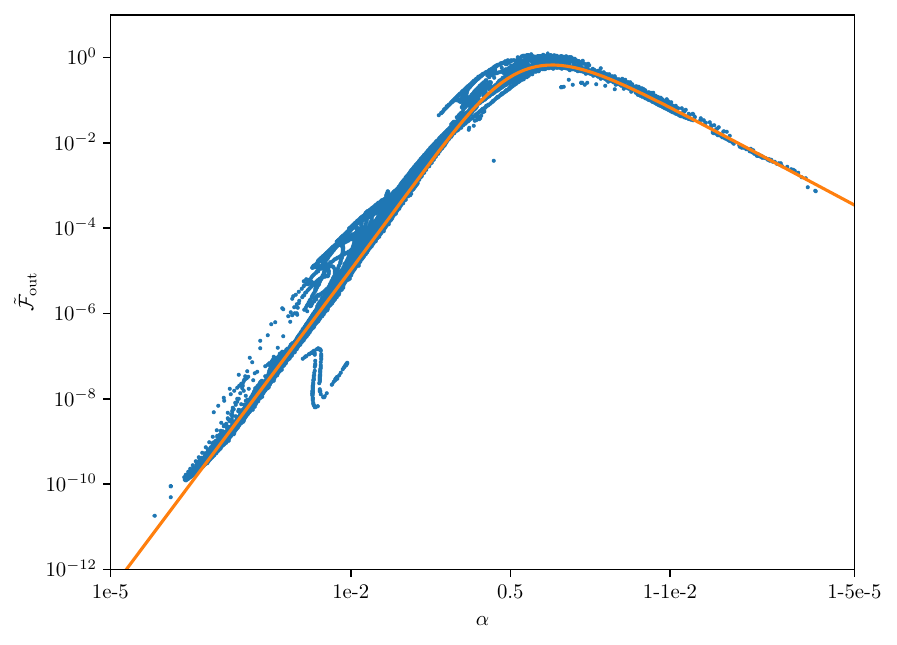}\vspace{-5pt}
        \includegraphics[width=\linewidth]{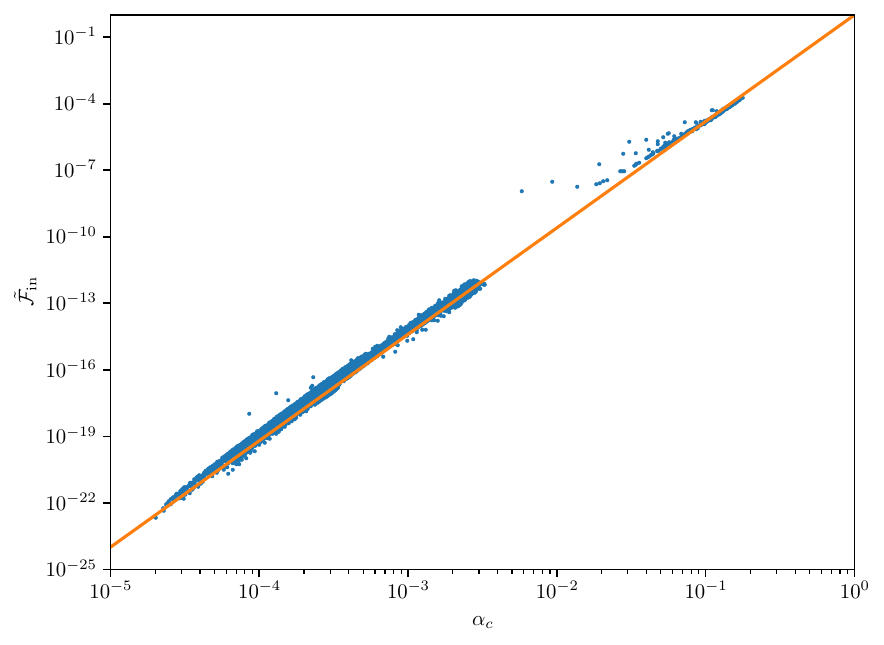}\vspace{-5pt}
        \caption{Unitless tidal forcing $\tilde{\mathcal{F}}$ for the envelope (top) and core (bottom) as a function of their respective radius aspect ratios. In blue all datapoints for all stellar evolutionary models computed in this work, and in orange the analytical approximation (Eq.~\ref{eq:tidalForcingApprox}).}\label{fig:tidalForcing}
    \end{figure}

    The unitless tidal forcing $\tilde{\mathcal{F}}$ is a complicated integral, but for all the models computed in this work (1 to 4 M$_\odot$), it is shown in Fig.~\ref{fig:tidalForcing} as a function of $\alpha = r_\text{in}/R_\star$. For high values of $\alpha$ (i.e. when the convective envelope is very thin), the tidal forcing decreases as $X$ does not have the room to grow. Gowing towards low values of $\alpha$ (i.e. when the convective envelope is very thick), the tidal forcing also decreases as $X$ targets the inner part of the integral, where $x$ is small. This results in a peak in the tidal forcing around $\alpha \approx 0.4$, where the convective envelope is neither too thin nor too thick. For an order of magnitude estimate, the dimentionless tidal forcing can be approximated as (a rough fit by eye):

    \begin{align}\label{eq:tidalForcingApprox}
        \tilde{\mathcal{F}}_\text{out} &= \alpha^{\frac{5}{2}}(1-\alpha)(1+6\alpha)\ ,\\
        \tilde{\mathcal{F}}_\text{in} &= \alpha_c^{4.8}\ .
    \end{align}

    This approximation is also shown in Fig.~\ref{fig:tidalForcing}, and captures the overall behaviour of the tidal forcing quite well.

\section{Froude number at the radiative-convective interface}\label{app:FroudeNumber}
\begin{figure}
    \centering
    \includegraphics[width=\linewidth]{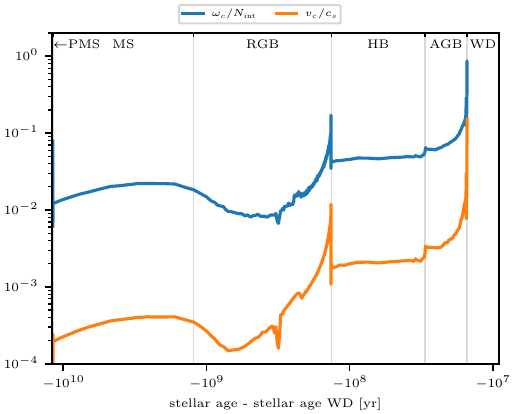}
    \caption{Froude number at the convective-radiative interface as a function of stellar age for a $M_\mathrm{ZAMS} = 1$ M$_\odot$ star. The Froude number computed as $\omega_c / N_{int}$ is shown in blue, and the Froude number approximated as $v_c / c_s$ is shown in orange. Stellar evolutionary phases (PMS to WD) are indicated.}\label{fig:Froude}
\end{figure}

\begin{figure*}
    \centering
    \includegraphics[width=\linewidth]{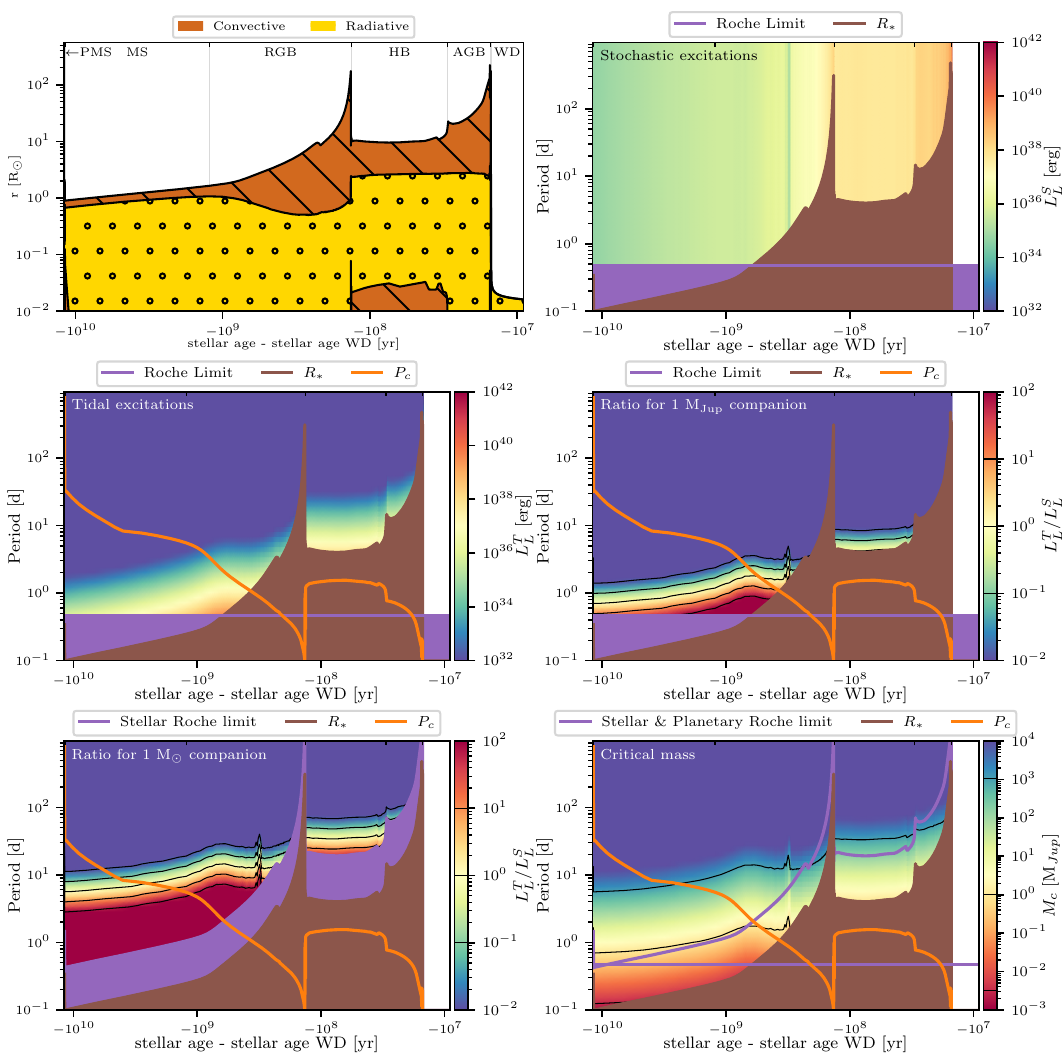}
    \caption{Same as Fig. \ref{fig:M20L}, but using the Froude number approximated as $v_c / c_s$ to compute the angular momentum luminosity carried by stochastically excited gravity waves.}\label{fig:M20L_cs}
\end{figure*}

    The Froude number at the convective-radiative interface is an important parameter in the calculation of the angular momentum luminosity carried by stochastically excited gravity waves (see Eq.~\ref{eq:LLS}). It is either computed as $\omega_c / N_{int}$ (in our approach following \citealp{Press1981}) or approximated as $v_c / c_s$ as proposed by \cite{Fuller2014}, where $c_s$ is the sound speed at the interface, because of the uncertainties on the stratification in the convective penetration region \citep[e.g.][]{Zahn1991,Pincon2016}. Their respective values are shown in Fig. \ref{fig:Froude} for the 1 M$_\odot$ stellar evolutionary model. The two definitions give differences up to 2 orders of magnitude. In the article, we can thus consider the Froude number computed as $\omega_c/N_{int}$ as proposed here as a maximum value scenario while the Froude number computed as $v_c/c_s$ is a minimal value scenario. When using the Froude number computed as $v_c / c_s$, the angular momentum luminosity carried by stochastically excited gravity waves will be lower by up to 2 orders of magnitude (can be seen in Fig. \ref{fig:M20L_cs}). Although this is a significant difference, the overall region of the parameter space where the tidal excitation starts to compete with the stochastic excitation will however not change significantly, as the dependance on the orbital period of the tidal excitation is so strong and our conclusions should be robust.

\section{Energy luminosities}\label{app:EnergyLuminosities}
\begin{figure*}
    \centering
    \includegraphics[width=0.95\linewidth]{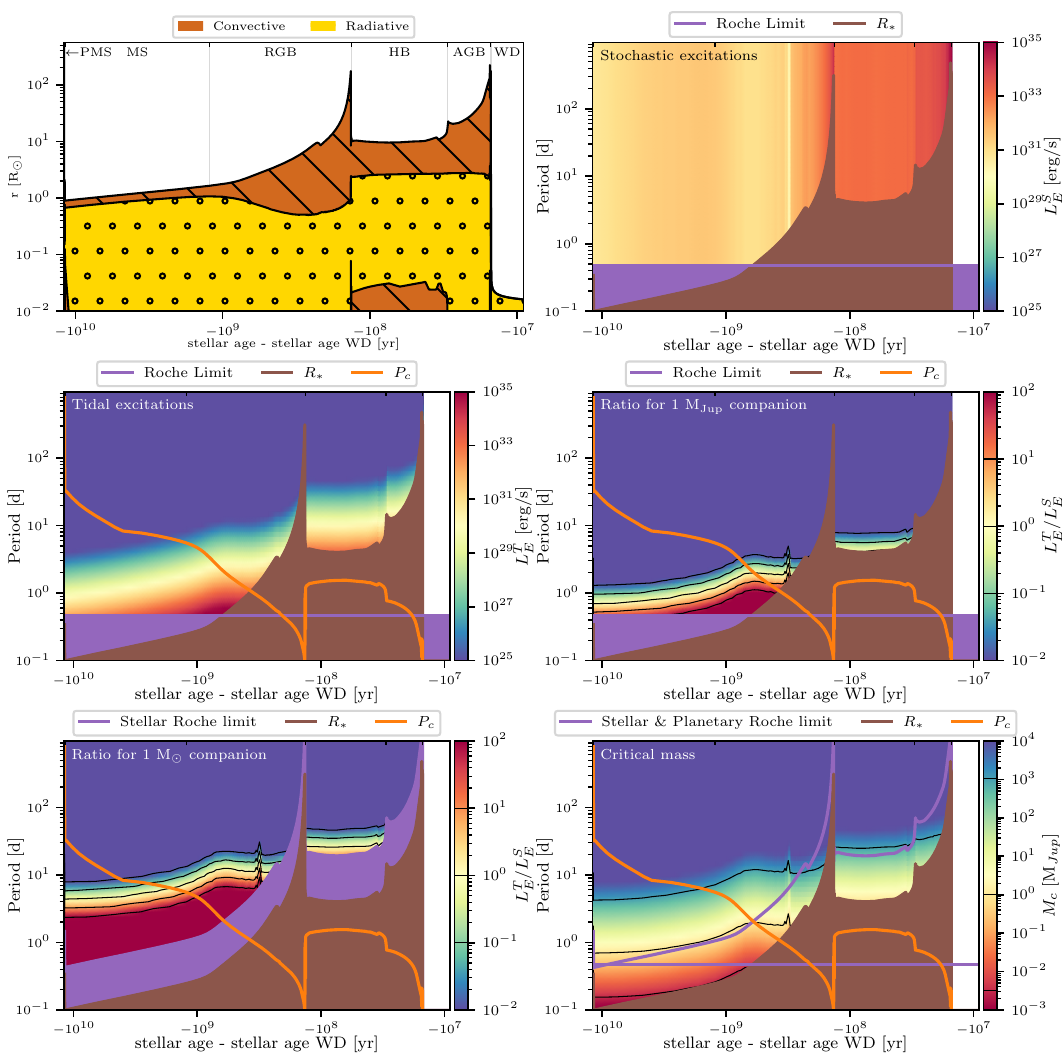}
    \caption{Same as Fig. \ref{fig:M20L}, but for the energy luminosities and for a 1 M$_\odot$ star.}\label{fig:M10E}
\end{figure*}

\begin{figure*}
    \centering
    \includegraphics[width=0.95\linewidth]{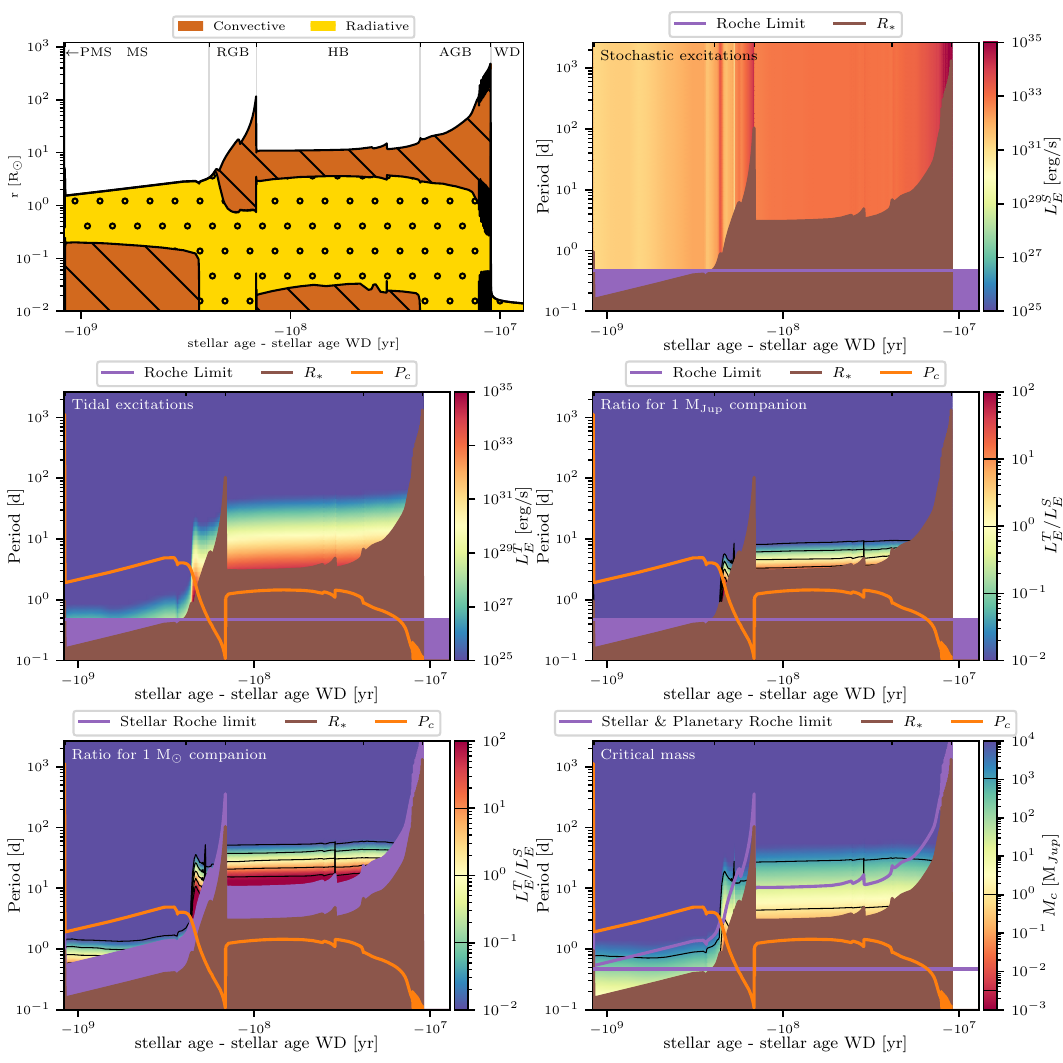}
    \caption{Same as Fig. \ref{fig:M20L}, but for the energy luminosities.}\label{fig:M20E}
\end{figure*}

    In Figs.~\ref{fig:M10E} and \ref{fig:M20E} the internal structure and energy luminosities carried by both stochastically and tidally excited gravity waves for the 1 M$_\odot$ and 2 M$_\odot$ stellar evolutionary models respectively are shown.

\end{document}